\documentclass[]{spie}  %>>> use for US letter paper
\newif\ifpdf
\ifx\pdfoutput\undefined
\pdffalse % we are not running PDFLaTeX
\else
\pdfoutput=1 % we are running PDFLaTeX
\pdftrue \fi
\ifpdf
\usepackage[pdftex]{graphicx}
\usepackage[pdftex]{color}
\DeclareGraphicsExtensions{.jpg,.png,.pdf} \else
\usepackage[dvips]{graphicx}
\usepackage[dvips]{color}
\DeclareGraphicsExtensions{.eps} \fi
\title{Stored light and EIT at high optical depths}
\author{
  M.\ Klein\supit{a,b},
  Y.\ Xiao\supit{a},
  M.\ Hohensee\supit{a,b},
  D.\ F.\ Phillips\supit{a},
  and
  R.\ L.\ Walsworth\supit{a,b} \skiplinehalf
  \supit{a}Harvard-Smithsonian Center for Astrophysics, Cambridge, MA,
  02138 USA \\
  \supit{b}Department of Physics, Harvard University, Cambridge, MA, 02138
  USA \\
}

\date{\today}
\begin{document}
  \maketitle

%%%%%%%%%%%%%%%%%%%%%%%%%%%%%%%%%%%%%%%%%%%%%%%%%%%%%%%%%%%%%

\begin{abstract}
  We report a preliminary experimental study of EIT and stored light
  in the high optical depth regime. In particular, we characterize two
  ways to mitigate radiation trapping, a decoherence mechanism at high
  atomic density: nitrogen as buffer gas, and a long, narrow cell
  geometry. Initial results show the promise of both approaches in
  minimizing radiation trapping, but also reveal problems such as
  optical pumping into trapped end-states. We also observe distortion
  in EIT lineshapes at high optical depth, a result of interference
  from four-wave mixing. Experimental results are in good qualitative
  agreement with theoretical predictions.
\end{abstract}


\maketitle

\section{Introduction}\label{intro}

Using electromagnetically-induced transparency (EIT) techniques to
store light for quantum communication
applications~\cite{LukinColloquium} is a topic of great current
interest. Proof-of-principle experiments in both cold and warm atomic
systems have successfully demonstrated reversible storage of photon
states~\cite{Caspar,Kimble-Store,Kuzmich-Store}, single-photon
EIT~\cite{EisamanNature}, and the application of stored light
techniques in quantum repeaters to increase the efficiency of long
distance quantum
communication~\cite{DLCZ,KimbleQR,KuzmichQR,PanJWQR}. Despite this
progress, low efficiency of storage and retrieval remains an obstacle
on the way to successful long distance quantum communication
applications~\cite{PanJWQR}.
% Among the many quantum memory candidates (which include cold atom
% ensembles, single atoms in cavities and ensemble-cavity hybrid
% systems),
Here, we investigate techniques towards improved efficiency in the
classical regime for  atomic vapor cell systems.
% remain interesting for their relative simplicity in construction,
% simpler energy level structure, clean magnetic field environments,
% and potential to scale.  [Is this last point true?].

The complete and reversible transfer between photon states and atomic
ensemble spin states relies on strong light-atom coupling. Large
coupling requires a correspondingly large optical depth. Theory based
three-level atom models reveal that stored-light efficiency tends
towards 100\% provided that the ground state decay rate is
zero~\cite{GorshkovLong}. In realistic atomic systems, decoherence
arises from mechanisms such as collisions with buffer gas atoms and
cell walls, diffusion out of the laser beam, and residual magnetic
field gradients.  When non-zero coherence decay is taken into
account, the storage efficiency initially improves with increasing
optical depth, but then falls off due to the increasing dominance of
absorption associated with the coherence decay. Such a trend has been
observed experimentally~\cite{SPIE2008,IrinaPRA}, but the efficiency
peaks at a much smaller optical depth than predicted, indicating that
additional loss mechanisms at large optical depth are missing from
the straightforward model.

Additional decoherence mechanisms include spin exchange, radiation
trapping and unwanted four-wave mixing. The presence of radiation
trapping at large optical depth in atomic vapor has been well
established~\cite{RT,SPIE2008}.
%, but its influence on stored light has not been fully explored.
Here, we present progress in our investigations of the effects of
radiation trapping on slow and stored light.
%Our previous study~\cite{our SPIE} based on fluorescence
%rise time detection and one-photon absorption linewidth verified that
%radiation trapping plays an important role in vapor cell systems at
%high optical depth. In the present paper,
We report preliminary
results on two potential methods to mitigate radiation trapping,
namely, by using a nitrogen buffer gas and by using a narrow cell. We
also present preliminary results on four-wave mixing effects on EIT.
Both the static EIT spectra and the stored light efficiency are
measured as diagnostic tools.

\section{Experimental Setup}\label{ExpSetup}

%Mention that the stored light efficiency was all achieved by using the iteration scheme.

%%% RUBIDIUM LEVELS %%%

\begin{figure}
  \centering{\includegraphics[width=0.6\columnwidth]{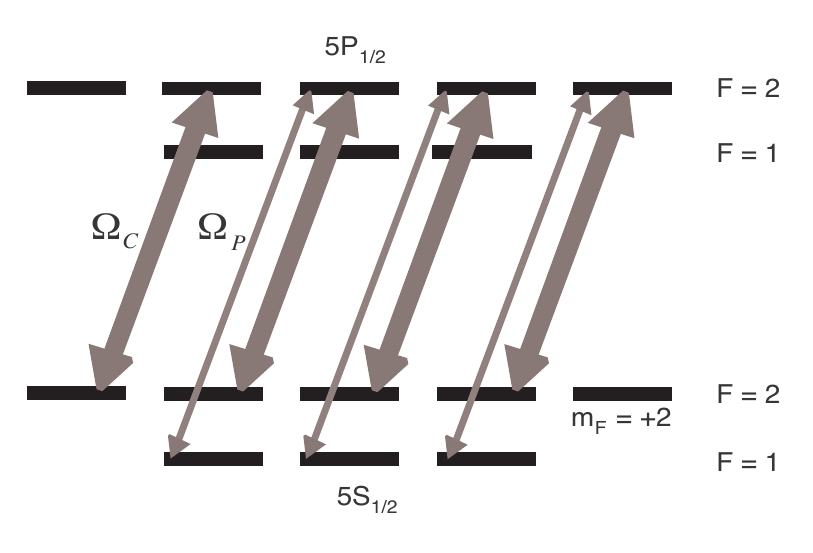}}
  \caption{
The D${}_{1}$ structure of ${}^{87}$Rb atomic energy levels and
coupling fields used in a hyperfine $\Lambda$-scheme for EIT and slow
and stored light.  Both light fields are $\sigma^{+}$-polarized, with
the strong control and weak probe fields  denoted by their respective
Rabi frequencies $\Omega_{C}$ and $\Omega_{P}$.  The $|F,m_F\rangle =
|2,2\rangle$ Zeeman sublevel acts as a trapped state, where atoms
optically pumped into the state cease participating in the EIT
process.
}
 \label{RbLevels}
\end{figure}

%%% EXPERIMENTAL SETUP %%%

\begin{figure}
  \centering{\includegraphics[width=0.6\columnwidth]{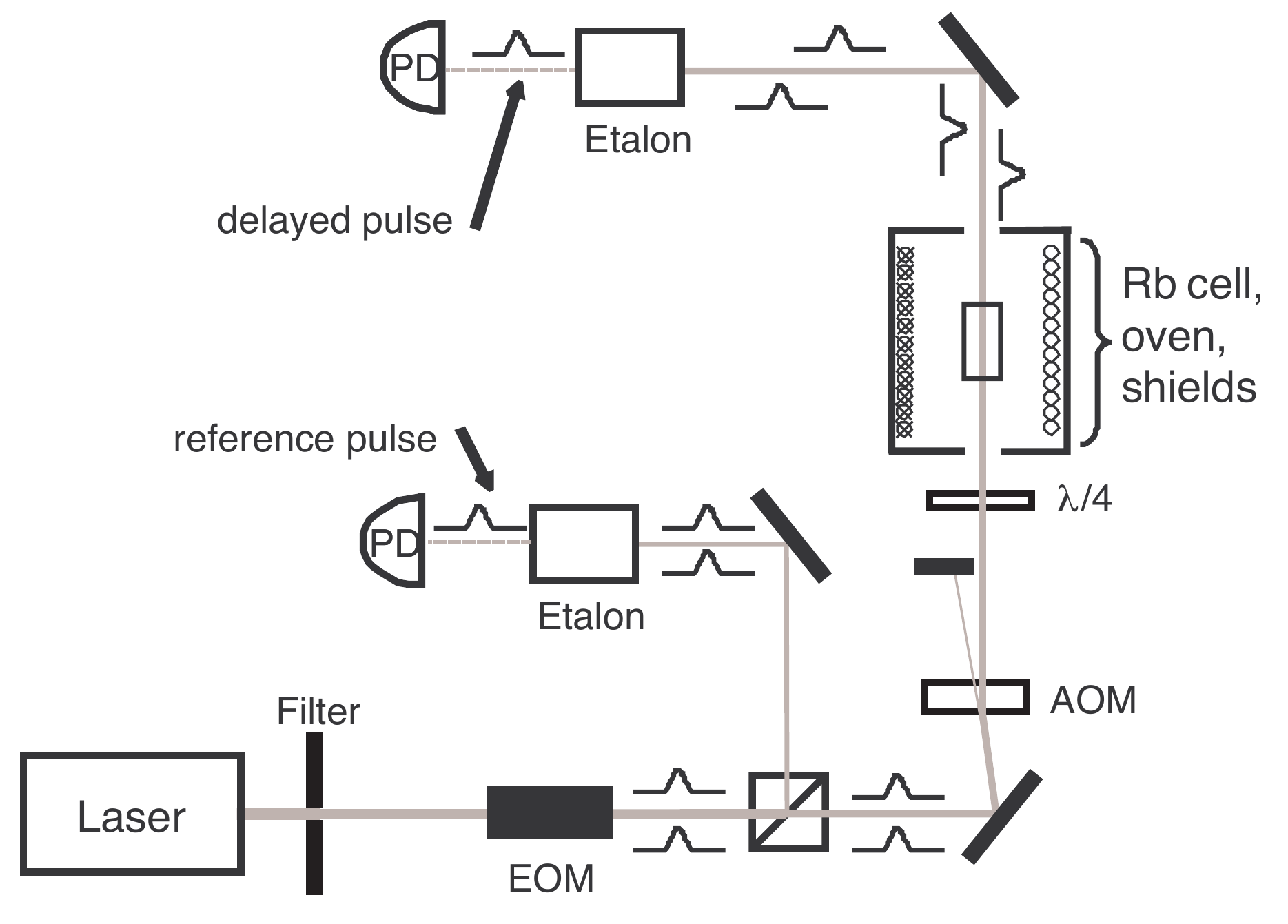}}
  \caption{ Apparatus used for EIT and stored light measurements in Rb
vapor cells.
%The laser output is split into a reference beam and main
%beam.
Electro-optic (EOM) and acousto-optic (AOM) modulators temporally
shape the probe and control fields which enter a rubidium (Rb) vapor
cell housed inside an oven with solenoid, and magnetic
shielding. Etalons are used to pass only the probe light onto
photodetectors (PD).  \emph{See text for details.} }
\label{Apparatus}
\end{figure}

We used the $D_{1}$ transition in a warm ${}^{87}$Rb vapor cell to
perform EIT and stored light experiments. The full energy level
structure and coupling fields are shown in Fig.\ \ref{RbLevels}: a
strong control field is resonant with the $F=2\rightarrow
F^{\prime}=2$ transition and a weak probe field with $F=1\rightarrow
F^{\prime}=2$. The atomic medium was in a low magnetic field
environment allowing $m=-1$, $m=0$, and $m=+1$ Zeeman levels to
participate in EIT processes (Fig.~\ref{RbLevels}).
%The signal field is constant for static EIT measurements and pulsed for slow and stored light.

The optical fields were generated by an external cavity diode laser
tuned to $\lambda=795$~nm, amplified by a tapered amplifier, and
spatially filtered through a pinhole as shown in
Fig.~\ref{Apparatus}. The amplified laser power
%output of the amplifier was about 100mW, and
after the pinhole was $70$~mW. A 6.8 GHz resonant electro-optic
modulator (EOM), driven with $\sim1$~W of rf power, phase modulated
the laser light at the ground state hyperfine splitting
($\sim6.835$~GHz); the carrier acted as the control field, and the
$+1$ sideband as the probe field, with a maximum probe to control
intensity ratio of $4\%$. The sideband amplitude was varied to shape
the desired probe pulse for stored light measurements.  The
EOM drive frequency was varied to change the two-photon detuning
$\delta$ of the EIT transition for measuring its lineshape.
For stored light measurements, both the drive and the probe fields
were quickly turned off by the AOM while the slow pulse traveled
through the vapor cell, storing the pulse in the medium. Then the
control field was turned on after a storage interval.

After an acousto-optic modulator (AOM) set the intensity of the light,
a quarter-wave plate ($\lambda/4$) converted it from linear to
circular polarization. The collimated beam entering the cell had a
diameter of approximately $6.8$~mm.  At the Rb cell output, a
temperature-stabilized etalon (free spectral range $20$~GHz and
extinction ratio of $\sim200$) was used to filter the control field
light and transmit only the probe light onto a photodetector (PD).
An etalon was used
%as opposed to other detection schemes such as beat note
%detection~\cite{our SPIE}
to attain the high bandwidth needed for detection of short pulses.
The transmission of the $-1$ EOM sideband (with $\delta$ set off resonance),
%being $\sim6$~GHz away from any Rb transition frequency,
was used as a reference for determining the off resonant transmission
level of the probe. With the carrier detuned $\sim13$~GHz below the
probe transition frequency $1\rightarrow2$, the $-1$ sideband is
$20$~GHz detuned, and thus transmitted by the next etalon resonance.
%The -1 sideband as a reference works because we sweep the laser while looking at the
%output.  When the carrier is ~13 GHz below the 1-->2 transition, the
%-1 sideband is 20 GHz away, which is the free spectral range.)
A small amount of input light was picked off earlier in the beam path
and sent to a separate temperature-stabilized etalon, and this signal
was used as a reference for slow light time delay measurements; slow
light delay was defined as the difference in peak arrival times
between the reference and output probe pulses.

The Rb vapor cell was housed inside a plastic oven, which was heated by blown
warm air.  The cell temperature was varied between $39$ and
$79~^{\circ}$C, or atomic number density between $3.8\times10^{10}$
and $1.1\times10^{12}$ cm${}^{-3}$.  Three layers of cylindrical
high-permeability shielding surrounded the oven to screen out stray
laboratory magnetic fields, and a solenoid inside was used to cancel
any small constant background field.
%
%As will be further described in sections below,
Three Rb vapor cells were used in our experiments; all three contained
isotopically enriched ${}^{87}$Rb and were filled with buffer gas to
confine atoms and extend their coherence life times. Two cells had
length $L=15$~cm and diameter $D=1.2$~cm, one filled with $40$~torr of
Ne buffer gas and the other with $25$~torr of N$_2$.  The third also
contained $40$~torr of Ne, and had dimensions $L=7.5$~cm and
$D=2.5$~cm.  The unusually long aspect ratio was used to combat the
effects of radiation trapping, as described below.  These cell choices
allowed us to directly investigate the role of different buffer gases
(by comparing the two long cells), and the role of cell geometry (by
comparing the two Ne buffer gas cells).

%We had two long cells with nitrogen and neon buffer gas respectively, and one short cell with neon buffer gas.

%Details on cell geometry and buffer gas pressure will be presented when the correspondent experiment is described.
%I disagree.  It takes little space to just give the dimensions here and cut this sentence.  It doesn't hurt
% to repeat it later if we need to.
%

%All reported stored efficiency here is obtained by an iterative procedure for finding the optimal probe light pulse shape to maximize efficiency at a ceratin cell temperature and laser power.

\section{Comparison of neon and nitrogen buffer gas vapor cells}

%%% N2 vs Ne BUFFER GAS %%%
\begin{figure}
  \centering{\includegraphics[width=1.0\columnwidth]{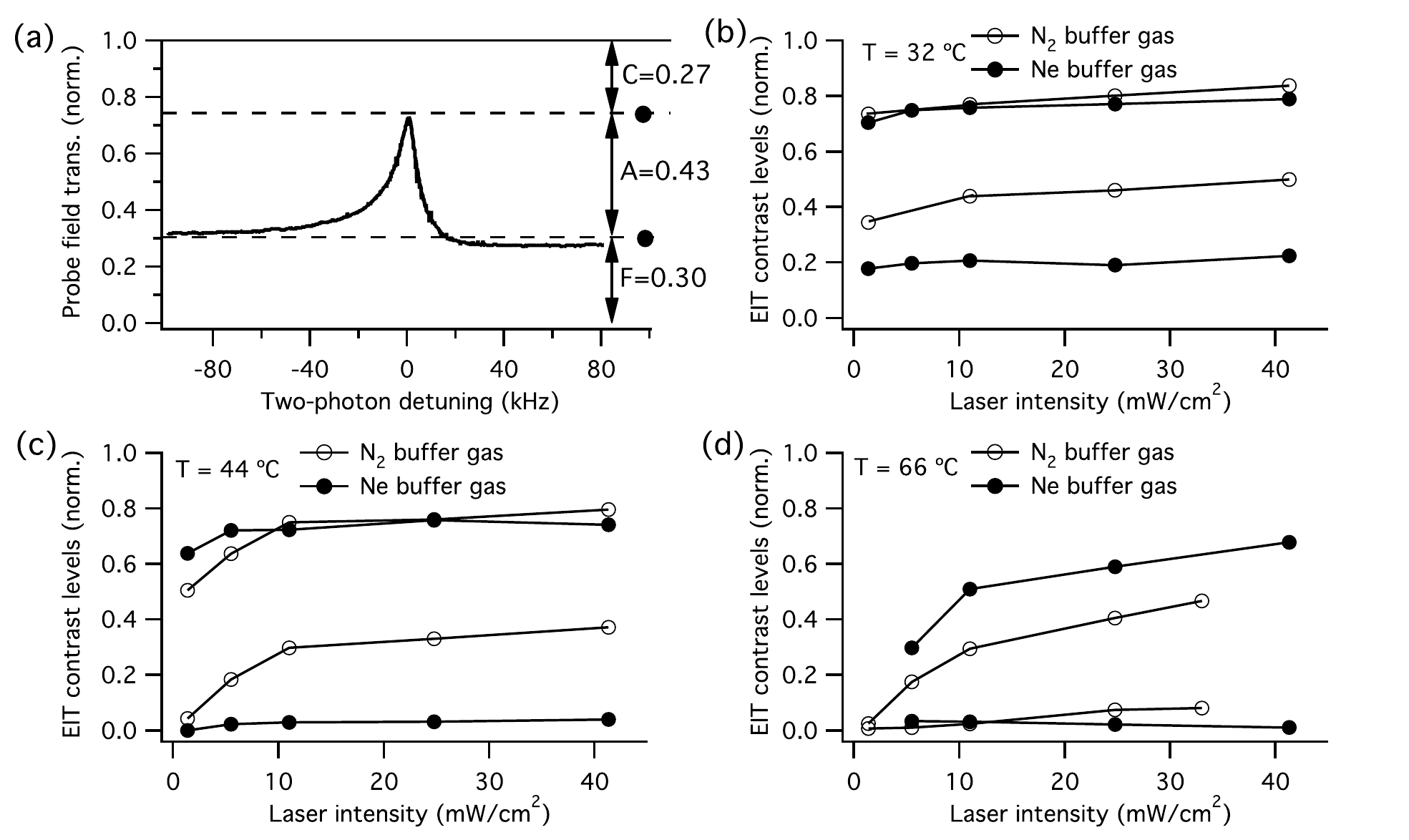}}
  \caption{
Neon and nitrogen buffer gas comparison, demonstrating reduced
effective optical depth in N$_{2}$.  (a) Typical EIT spectrum,
normalized to the full off-resonant probe 
transmission level.  The contrast levels are labeled by $F$, the
off-resonant transmission level, which is a measure of the system's
optical depth; $A$, the EIT amplitude; and $C$, the normalized difference
between full transmission (away from the Rb D${}_{1}$ transition) and the
peak of the EIT spectrum.  EIT resonance measured at $T=44\
{}^{\circ}$C and laser power $4$~mW. EIT contrast levels $F$ and $F+A$
vs.\ laser intensity for both Ne and N${}_2$ are shown for Rb cell
temperatures (b) $T = 32\ {}^{\circ}$C, (c) $T = 44\ {}^{\circ}$C, and
(d) $T = 66\ {}^{\circ}$C.  
}
 \label{N2vsNe}
\end{figure}
We compared the use of Ne and N$_2$ buffer gases in EIT
with the aim of mitigating radiation trapping in
stored light media.
Radiation trapping occurs when atoms absorb input light, then undergo
spontaneous emission; these emitted photons can then be re-absorbed
and re-emitted multiple times before escaping an optically-thick
atomic medium. Since spontaneously-emitted photons have random
polarization, re-absorbed photons act to equilibrate the population
distribution among different Zeeman sublevels, reducing the optical
pumping efficiency and coherence lifetime.  Radiation trapping can
significantly impact the polarization of an EIT
medium~\cite{RT,SPIE2008} and thus potentially the efficiency of
stored light.

To reduce the negative effects of radiation trapping, we tested
nitrogen as buffer gas, as nitrogen de-excites the Rb atoms without
scattering photons by transferring the energy to vibrational states in
the N$_2$ molecules~\cite{RT}. We note that many optical pumping
experiments have used nitrogen for this purpose but these studies use
much higher nitrogen pressure than we report here in EIT
experiments~\cite{Nitrogen}. We used two identical $^{87}$Rb vapor
cells, one with $40$~torr of Ne, and the other with $25$~torr of N$_2$
chosen such that the excited state pressure broadening ($\gamma$) was
the same in both cells.
The nominal optical pumping rate, $\Omega_C^2/\gamma$, was then the
same in both cells for a given laser power.
% This is also the rate to establish ground state coherence.
Since radiation trapping depolarizes and decoheres the atomic ground
state, we used measurements of EIT (which is based on ground state
coherence) in the two cells to study the effect of radiation trapping.

%therefore keeping the former the same, while looking at remaining coherence(via EIT measurements) gives clue on the decoherence rate.

%To test how an N$_2$ and Ne buffer gas cells compare, we measured EIT
%spectra;
EIT is sensitive to both optical depth and ground state decoherence,
and is thus a useful proxy for stored light
experiments~\cite{MasonOL}. In particular, the off-resonant floor of
an EIT spectrum is determined by the effective optical depth $d$:
$F=\exp{(-d)}$, where $d$ is proportional to the optical path length,
the atomic density, and the fraction of atoms in the relevant atomic
levels.  The peak transmission is determined by the control field Rabi
frequency, the effective optical depth $d$ and the ground state
coherence decay rate $\gamma_0$ : $ F+A=\exp{(-d \gamma
  \gamma_0/\Omega_C^2)}$.  Here, the floor ($F$) and amplitude ($A$)
are normalized to the off-resonant transmission level of the probe
light, as shown in Fig.\ \ref{N2vsNe}(a). In general, a lower floor
and a higher EIT transmission are desirable for stored light
experiments.

We observed lower EIT floors and higher peak transmissions in
neon buffer gas cells than N${}_2$ cells (Fig.~\ref{N2vsNe}).
%Our experiment results shown in Fig.\ \ref{N2vsNe}
%(b-d)indicate that the $\emph{neon}$ buffer gas cell yields better EIT: both a lower
%floor and a higher peak transmission than in the nitrogen cell.
%, more noticeable in the high laser intensity and higher temperature respectively.
%
Therefore, we expect reduced stored-light performance from the N${}_2$
buffer gas cell in comparison to the neon cell.
We believe the increased EIT floor in the N${}_2$ cell was
predominantly due to increased population in the $|F=2,
m_{F}=2\rangle$ end state.
%
%can be mostly explained by the trapped
The $\sigma^{+}$ polarization of the control and probe light led to
optical pumping into this state, which was then inaccessible to the
EIT process, effectively reducing the optical depth and raising the
EIT floor. As seen in Fig.\ \ref{N2vsNe}, the floor level increased
with laser intensity in both cells because more atoms were pumped to
the end state at high laser intensity. The higher EIT floor in the
N$_2$ buffer gas cell was indicative of more efficient optical pumping
to the trapped state, and poorer performance for slow and stored
light.

Since radiation trapping  reduces
optical pumping efficiency, the higher trapped state population in the N$_2$ cell
may indicate that this cell had less radiation trapping.
However,
%reduced radiation trapping is not the only
%interpretation for a higher floor in N$_2$ cell. The different ways
%that
the population distribution as Rb de-excites to the ground state in
the presence of different buffer gases may also play a role. We
carried out a full 16-level density matrix simulation of both cells in
the optically thin regime, and found that if the Rb atoms are quenched
to the ground state with the same $m_F$ state before excited state
mixing occurs (as in the N$_2$ cell) there is a higher population in
the trapped state. However, if Rb is allowed to de-excite to all
allowed ground states (as in the Ne cell, since the excited states are
uniformly mixed before spontaneous emission), then there is less
population in the trapped state. Detailed modelling of the population
distribution after optical pumping requires branching ratios of Rb
de-excitation in the presence of various buffer gases and further
study will be required.

%%% [I don't think this paragraph is true!!! I don't think that the
%%% low power line width is small enough for diffusion out of the 6.8
%%% mm beam to play a significant role in determining the coherence
%%% time.]
The N$_2$ buffer gas cell also showed lower peak EIT transmission
(Fig.~\ref{N2vsNe}) due to the larger ground state decoherence rate in
the nitrogen cell.
%, which is determined by the
%interaction time (diffusion time). Rb atoms diffuse faster through
%N$_2$ than through Ne, and the pressure of N$_2$ is less, which
%combines to give a shorter interaction time in nitrogen [is it four
%times different? need to find the numbers of diffusion constant].
We also measured the ground state coherence decay rate by the 1/e time
of stored light efficiency vs.\ storage time, and found a decoherence
rate that was four times greater in the nitrogen cell.  Shorter
atom-light interaction time (the diffusion time) in the nitrogen cell
contributed to this increase in decoherence.
The N$_2$ buffer gas did not provide better EIT than Ne buffer gas,
due to increased pumping of Rb atoms into the trap state. While
further work is needed to understand the details of Rb de-excitation,
%process
%of Rb in nitrogen and to solve the trapped state problem, it is also
other techniques must also be developed to reduce the effects of
radiation trapping in stored light vapor cells.

\section{Narrow vapor cell geometry}

%%% CELL GEOMETRY AND RADIATION TRAPPING %%%

\begin{figure}
  \centering{\includegraphics[width=1.0\columnwidth]{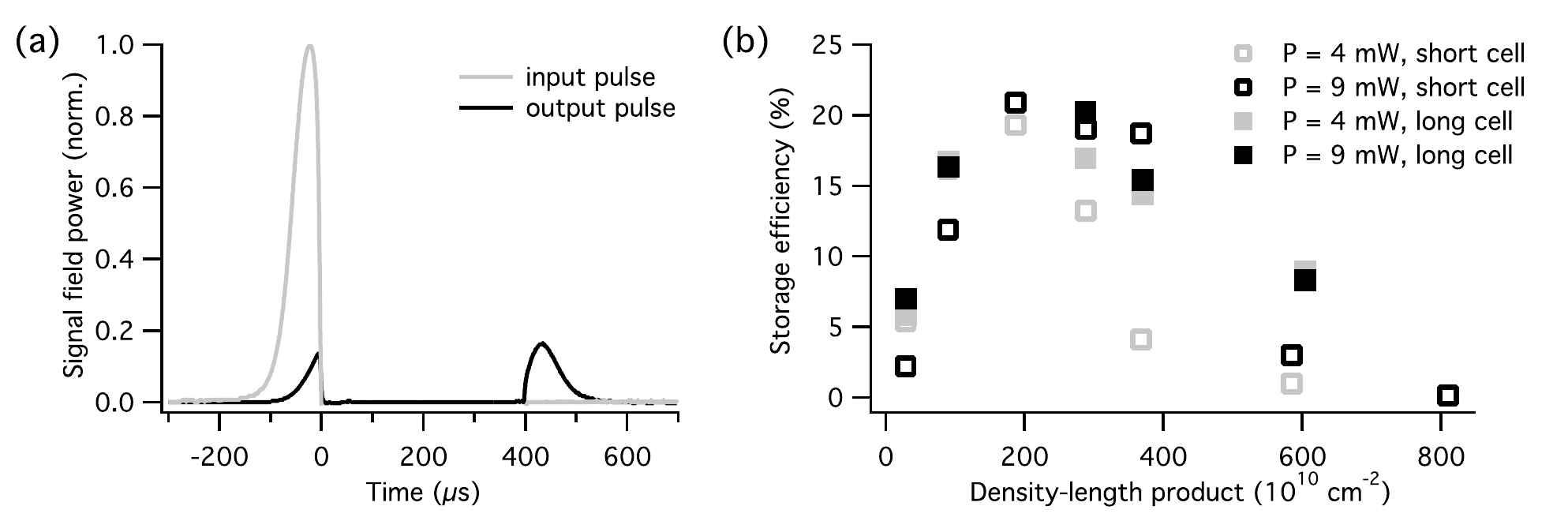}}
  \caption{ Stored light in Ne buffer gas cells, and the dependence on
    geometry.  (a) Example stored light measurement in the long,
    narrow $40$~torr Ne buffer gas cell.  The input reference pulse is
    normalized to the off-resonant transmission level for the probe
    field.  The output probe pulse consists of a leaked part, escaping
    before the control field is shut off, then a readout after a
    $400~\mu$s storage interval.  Leakage is present because the
    entire pulse is not spatially compressed enough to fit inside the
    cell.  $T=57~{}^{\circ}$C, and the laser beam diameter and power
    $d=6.8$~mm and $P=9$~mW.  Efficiency is defined as the area of the
    output pulse divided by the area of the input pulse, here
    $20.3\%$.  (b) Geometry-dependence of stored light, using Ne
    buffer gas cells. At high optical depth, the short cell yields
    very low efficiency, while the long, narrow cell has significantly
    greater efficiency. We attribute the improved storage efficiency
    to a reduction in radiation trapping due to reduced transverse
    optical depth from both a smaller cell diameter and lower
    Rb density.}
 \label{Geometry}
\end{figure}

In addition to using nitrogen buffer gas to quench the excited state
in Rb atoms, radiation trapping can be mitigated by using a long,
narrow cell geometry.  A narrow cell allows fluorescence to escape the
cell in the transverse direction with fewer depolarizing interactions
with Rb atoms. When the transverse diameter of the cell is kept below
one optical depth of the Rb medium, most radiated photons escape
before undergoing additional interactions with the Rb vapor. (Such
aspect ratios were also investigated in a cold atom cloud, where
radiation trapping was reduced, enabling laser cooling to much higher
phase-space densities~\cite{Mukund}.)  A longer cell can reach the
same optical depth with a lower atomic density,
%needs a lower atomic density to reach the same optical depth as in shorter cell which should
allowing the vapor cell to be heated to lower temperatures and
reducing additional density-dependent effects
% including radiation trapping
% wave mixing,  <- that's total optical depth dependent not density dependent
such as spin-exchange collisions.

To test the effects of cell aspect ratio, we compared stored light
results in two $40$~torr Ne buffer gas cells with different
geometries: a longer cell with $D=1.2$~cm and $L=15$~cm, and a shorter
cell with $D=2.5$~cm and $L=7.5$~cm.
%of 15 cm long and 1.2 cm diameter, filled with 40 Torr neon, and
%compared the stored light performance with that in a 7.5 cm long and
%2.5 cm diameter cell filled with the same amount of neon.
%The aspect ratio in the thin cell is four times larger.
%Storage efficiency was maximized for each temperature and power using the
%scheme described in \cite{}, where the ideal pulse shape is determined
%iteratively.
Optimal stored-light efficiency was measured using an iterative
optimization process~\cite{GorshkovPRL,Expt-iteration} as a function
of optical depth for a fixed storage time of $\tau=400~\mu$s, at two
laser powers. Initial results (Fig.\ \ref{Geometry}) showed improved
efficiency in the long cell than the short cell.
%, storage efficiency drops more gradually with optical depth than in the short
%
At high optical depth, the long cell yielded much higher efficiency,
indicative of reduced radiation trapping in the high aspect ratio
cell. However, the theoretically expected larger overall
efficiency~\cite{GorshkovLong} was not achieved. We attribute this to
the presence of greater magnetic field inhomogeneity in the long
cell. The addition of higher order compensating coils will reduce this
inhomogeneity and may improve the storage efficiency.
Our measurements suggest that a high aspect-ratio cell is a route
towards improved efficiency.

\section{Effects of four-wave mixing}

%%% FOUR-WAVE MIXING %%%

\begin{figure}
  \centering{\includegraphics[width=1.0\columnwidth]{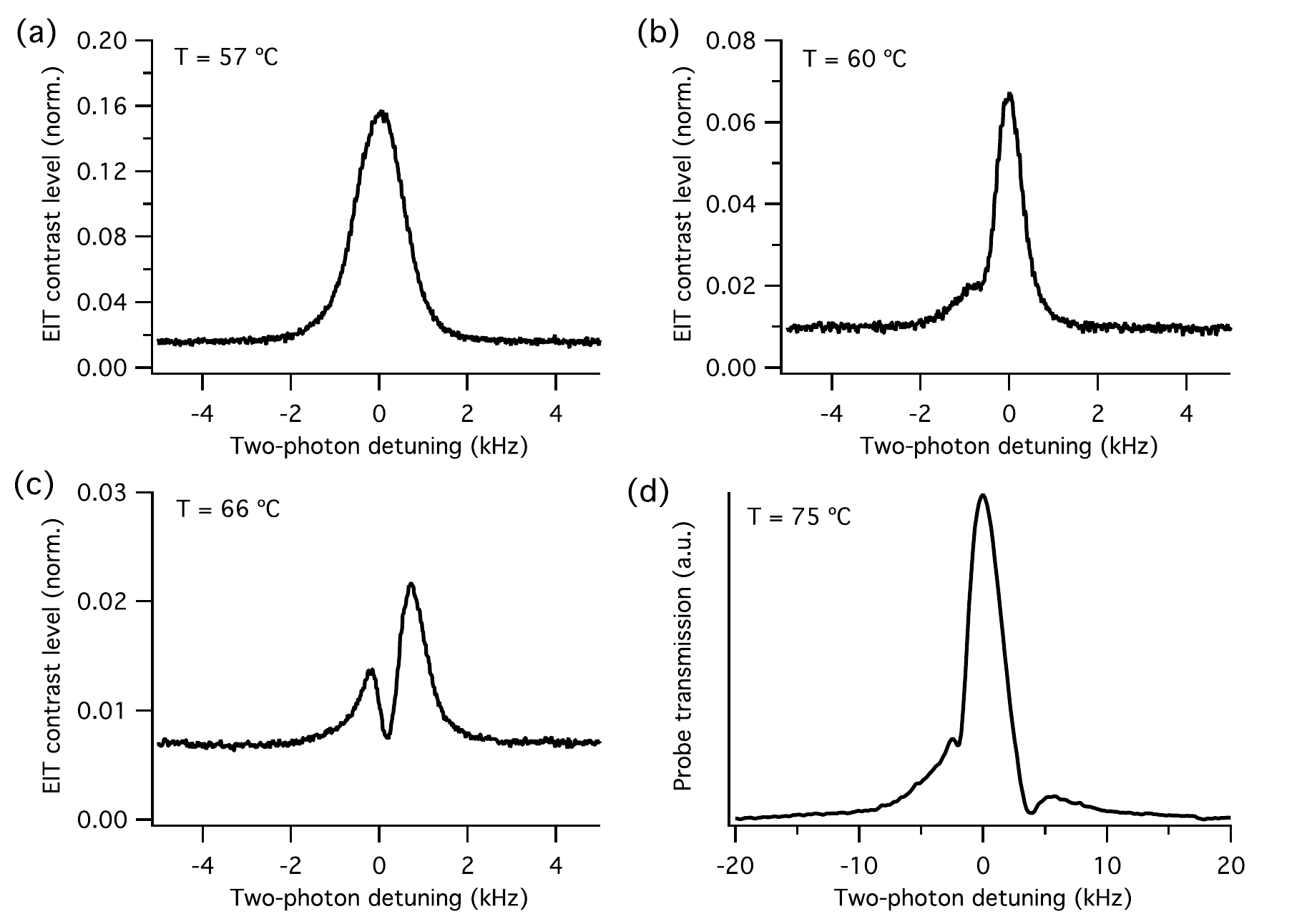}}
  \caption{ Additional peaks in EIT spectra due to four-wave
    mixing. (a-c) Using the long, narrow N$_2$ buffer gas cell, for a fixed
    input power of $P=500~\mu$W at respective temperatures of $T=57~{}^{\circ}$C, 
    $T=60~{}^{\circ}$C, and $T=66~{}^{\circ}$C.  (d) Using the $7.5$
    cm Ne buffer gas cell at $T=75~{}^{\circ}$C, with input power $P=8.8$~mW, we observe
    three peaks in the spectrum; relative size of the four-wave mixing peaks is diminished 
    at this much higher laser intensity.  }
 \label{4WM}
\end{figure}

Although electromagnetically induced transparency describes most of
the physics in the Rb medium coupled to two optical input fields,
% I don't know -- we just spend awhile discussing the important role
% of the trapped state, which is by definition not in the three-level model
% .fixed.
four-wave mixing plays an important role in the high-density limit
because its efficiency scales with the square of the density. The
control field can interact with the probe transition, generating an
anti-stokes field, and forming a four-wave-mixing cycle. Although the
drive field is detuned from the probe's resonance by $6.8$~GHz, EIT
ground state coherence can enhance the efficiency of nonlinear wave
mixing~\cite{Harris-FWM}. Furthermore, in our experiment, the unwanted
$-1$ sideband of the EOM coincides with the
(control-probe-control)-generated anti-stokes resonance and seeds
four-wave mixing.  Four-wave mixing coexisting with EIT has been
reported in several
experiments~\cite{IrinaPRA,LukinDensityNarrow,taohong}. For example,
EIT with multiple peaks (oscillations or fringes) caused by four-wave
mixing was observed in a high optical depth Rb sample cooled by buffer
gas~\cite{taohong}. Wave mixing can also cause gain in the weak
probe. We have directly observed greater than $100\%$ transmission of
the probe light in a miniature paraffin-coated
cell~\cite{small-coated-cell}. This gain induces noise which can
destroy quantum correlations in the photon pairs generated in such
systems~\cite{PanJWPRA}.

We studied EIT lineshapes for several different optical depths
(Fig.~\ref{4WM}). Since EIT lineshapes can be changed by adjusting the
one-photon detuning~\cite{Irina-EIA}, we set the laser to resonance
with $F=2$ excited state to simplify analysis.  However, the EIT
contrast improved when the laser was detuned from the $F=2$ resonance;
in particular, the stored light efficiency was higher when the laser
was tuned to the $F=1$ excited states than to the $F=2$, which we
attribute to an effectively lower optical depth from the reduced
coupling of the transition to the $F=1$ transition.
At large optical depths the EIT lineshape had multiple peaks or
fringes~\cite{LukinDensityNarrow,taohong} as shown in Fig.~\ref{4WM},
and consistent with theoretical
analysis~\cite{LukinDensityNarrow,taohong}. Fringes in the EIT
resonance reduce the useful transparency window and the stored light
efficiency.  The period of the fringes in an EIT lineshape
contaminated by four-wave mixing is $\sim\Omega_C^2/ \gamma d$, where
$d$ is the homogeneous optical depth, and the EIT linewidth is
$\Omega_C^2/ \gamma \sqrt{d}$~\cite{LukinDensityNarrow}. The fringe
period scales linearly with $d^{-1}$, reflecting the dispersive
characteristics of the system and the number of oscillations in the
EIT resonance scales as $\sqrt{d}$.
At high optical depth the efficiency of four-wave mixing is also
higher, increasing the fringe contrast. Therefore the contrast and
number of fringes in the EIT spectrum can be used as a diagnostic of
the optical depth and four-wave mixing in the system and set an upper
limit on the available optical depth for good stored light
performance in a vapor cell.
Further study is needed to evaluate the negative effect of four-wave
mixing on stored light efficiency (though the efficiency of the
retrieved anti-stokes field has been studied
previously~\cite{IrinaPRA}). Active elimination of four-wave mixing
via sophisticated polarization schemes have also been proposed
theoretically~\cite{Hohensee}.

\section{Conclusions}

We have reported a preliminary experimental study of EIT and stored
light aimed at understanding performance at high optical depth. We
compared N$_2$ and Ne buffer gases, and studied stored light efficiency
in long, narrow vapor cells. Initial results with both N$_2$ and long,
narrow cells show promise for reducing the effects of radiation
trapping.  Pumping of population into trapped end states needs to be
addressed before one can make effective use of an N$_2$ buffer gas
cell.  The long, narrow cell is also a promising route to high
efficiency with improved magnetic field compensation. We also observed
multiple peaks in EIT lineshapes, which became stronger at high
optical depth due to interference between EIT and a four-wave mixing
channels. Experimental results were in good qualitative agreement with
theoretical predictions.

We are grateful to I.\ Novikova for useful discussions. This work was
supported by ONR, DARPA, NSF, and the Smithsonian Institution.

\end{document}